\documentclass{hotnets15}

\usepackage{hyperref}
\usepackage{graphicx}
\usepackage{subfigure}
\usepackage{times}
\usepackage{balance}
\usepackage[font=small,format=plain,labelfont=bf,up,textfont=it,up]{caption}

\graphicspath{{figs/}}
\DeclareGraphicsExtensions{.png,.pdf}

\usepackage{color,url}

\newif\ifcomments
\commentstrue

\ifcomments
  \newcommand\note[2]{{\color{#1}\bf\em #2}}
  \newcommand\mort[1]{{\note{green}{mort: #1}}}
  \newcommand\nn[1]{{\note{blue}{N: #1}}}
  \newcommand\gareth[1]{{\note{red}{G: #1}}}
\else
  \newcommand\note[2]{}}
  \newcommand\mort[1]{}
  \newcommand\nn[1]{}
  \newcommand\gareth[1]{}
\fi

\newcommand{\one}{({\em i}\/)}
\newcommand{\two}{({\em ii}\/)}
\newcommand{\three}{({\em iii}\/)}
\newcommand{\four}{({\em iv}\/)}
\newcommand{\five}{({\em v}\/)}

\renewcommand\paragraph[1]{\vspace*{0.5cm}\noindent {\bf #1.}}

\renewcommand{\tabcolsep}{4pt}

\makeatletter
\def\url@leostyle{%
  \@ifundefined{selectfont}{\def\UrlFont{\sf}}{\def\UrlFont{\sffamily}}}
\makeatother
\makeatletter
\renewenvironment{thebibliography}[1]{%
  \begin{oldthebibliography}{#1}%
    \setlength{\parskip}{0ex}%
}%
{%
  \end{oldthebibliography}%
}
\makeatother

\title{Staggercast: Demand-Side Management for ISPs}

\author{Gareth Tyson,$^{\diamond}$ Nishanth Sastry,\textsuperscript{$\bullet$} Richard Mortier\textsuperscript{\textdaggerdbl} and Nick Feamster$^{\triangle}$ \\
	 $^{\diamond}$Queen Mary University of London,  \textsuperscript{$\bullet$}King's College London, \\
	 \vspace{6pt}
	  \textsuperscript{\textdaggerdbl}Cambridge University, $^{\triangle}$Princeton University \\
	 $^{\diamond}$gareth.tyson@qmul.ac.uk,  \textsuperscript{$\bullet$}nishanth.sastry@kcl.ac.uk, \\ \textsuperscript{\textdaggerdbl}richard.mortier@cl.cam.ac.uk, $^{\triangle}$feamster@cs.princeton.edu
}

\begin{document}
\maketitle

\begin{sloppypar}

\begin{abstract}
  The continuing expansion of Internet media consumption has increased
  traffic volumes, and hence congestion, on access links. In response, both mobile and wireline 
  ISPs must either increase capacity or perform 
  traffic engineering over existing resources. Unfortunately,
  provisioning timescales are long, the process is costly, and
  single-homing means operators cannot balance across the last
  mile. Inspired by energy and transport networks, we propose
  demand-side management of users to reduce the impact caused by
  consumption patterns out-pacing that of edge network provision. By
  directly affecting user behaviour through a range of incentives, our
  techniques enable resource management over shorter timescales than is
  possible in conventional networks. Using survey data from 100 participants we explore the feasibility of introducing the principles of demand-side management in today's
  networks.

%
%
\end{abstract}

\section{Introduction}
\label{sec:intro}

The Internet is now dominated by voluminous rich-media video and audio flows~\cite{LabovitzInternetPeering}. Delivering these flows is significantly more expensive for content providers due to strict constraints on bandwidth. At the same time, the size of rich-media flows means that a single content provider can overwhelm an entire access Internet Service Provider (ISP). For instance, ISPs noted significant increases in costs related to streaming for BBC iPlayer when it was first launched in the UK: in just four years, iPlayer rose to 6.4\% of all UK Internet traffic. In the US, Netflix streams dominate access networks, with a 32.9\% traffic share during peak hours~\cite{sandvine}.

These changes mean that Internet traffic, and particularly the traffic carried by ISPs serving the domestic market, is increasingly dominated by the desires and behaviours of end users. Unfortunately, while capacity upgrades and provisioning solutions are feasible in the network core, it is far more expensive to upgrade the geographically-dispersed access networks that reach domestic users. Availability of capacity is an even bigger challenge for mobile environments. Despite this, the various \emph{choices} made by these users rarely surface when considering practical solutions. Instead, the focus is on lower layers: Traffic peaks are treated as natural phenomena exhibited through increased packet rates on network interfaces. This has led to ``net neutrality'' issues, whereby ISPs attempt to shape certain types of traffic. However, to the everyday person, this concept remains opaque, compounding concerns vocalised by campaigners~\cite{cheng2011debate}.

As exemplified by this ongoing tussle over net neutrality, content providers and ISPs each consider the other culpable. Neither consider that it is the demands of \emph{users} that is the underlying cause of their problems. Our argument is that, in addition to managing the \emph{effects} of user behaviour (i.e., managing traffic surges and other packet-level phenomenon), ISPs should also explore ways to alter the {\em cause} of these phenomena in the first place, by affecting user behaviour. We argue this is a positive step and should be encapsulated in a framework that transparently communicates all forms of potential traffic manipulation to the user. By including users in this control loop, many novel forms of traffic engineering become possible; importantly, these can be done with user consent, rather than in the current opaque fashion.

Thus, this paper proposes that ISPs introduce incentives to modify user behaviour, shifting users away from undesirable (peak-time) activity. Although early work (e.g., time dependent pricing~\cite{Sen12}) has explored adapting user behaviour, we make the case for integrating several novel approaches into a single unified architecture driven by ISP goals. Taking inspiration from \emph{demand-side management} (DSM) techniques in energy and transport networks, which aim to reduce peak-to-mean ratios by shifting load, we examine the potential of proactively adapting user behaviour through a range of targeted incentives~(\S\ref{sec:dsm}). We then propose a simple strawman architecture, \emph{Staggercast}, allowing ISPs to incentivise behavioural change in  pursuit of their own goals~(\S\ref{sec:managing}). Unlike existing forms of traffic shaping, we argue that a key focus should be transparency, allowing users to fully understand the needs of their ISP. This would offer an effective middle ground between the two extremes of the net neutrality debate. Specifically, we propose DSM over three dimensions: \one~\emph{time}, where users could be encouraged to shift demands away from peak periods; \two~\emph{space}, where users are encouraged to utilise certain (lightly loaded) parts of the network; and \three~\emph{content}, where users could be encouraged to consume content with lower network costs. To explore these ideas we analyse user willingness to engage with Staggercast using a questionnaire~(\S\ref{sec:survey}). We then discuss the relationship of Staggercast to prior incentive schemes~(\S\ref{sec:related}), and conclude with a discussion of concerns and research challenges~(\S\ref{sec:summary}).

\section{The Case for DSM}
\label{sec:dsm}

Demand-side management (DSM) is the practice of shaping user behaviour to better satisfy some goal. Although rudimentary demand pricing has been introduced in some countries already, this is generally the limit in terms of commercial ISPs. For wireline broadband, indiscriminate ``all you can eat'' contracts remain the norm (at least in many developed countries). Mobile networks are moving in a similar direction. We begin with two questions: \one~How could specific user activities be ``managed'' to better satisfy ISP goals? and \two~What incentives could drive users towards accepting such demand management?


\subsection{Internet DSM Techniques}

Most user activities affect the network to some extent, whether the user is sending an email, streaming a video, or posting a tweet. An ISP could apply DSM to all of these traffic flows to better satisfy their own objectives, such as reducing transit costs or alleviating the load on certain areas of the network. Given that the dominant cause of network load is video, we focus on three types of behavioural change that involve \emph{shifting} that specific demand:

\one~{\bf Time shifting}. An obvious cause of high peak loads is temporally correlated user activity. In these situations, DSM could encourage users to change the \emph{time} that they use the Internet to reduce the size of these peaks. Although much work exists on time shifting~\cite{sen2013price}, there are many variations that can be explored. For instance, users could be asked to pre-fetch content overnight or to temporarily use less heavyweight applications. This would be particularly beneficial for mobile providers, who could ideally offload traffic onto the wireline network, e.g.,~by pre-fetching content in the morning. Less dramatically, ISPs might dynamically introduce ``filler'' content (such as advertisements or extended introductions) into streams from local sources, effectively staggering requests for content without the users' knowledge. Although not helpful for heavily congested edge links, this could certainly help reduce transit peaks. Offsetting consumption patterns in these ways, even for live events, might also benefit energy networks  by de-synchronising start, break, and finish times for viewers (energy providers could perhaps even incentivise ISPs to achieve this). During popular real-time events, ISPs could also ask that users delay other tasks (e.g. syncing Dropbox) until the event has finished.

\two~{\bf Space shifting}. In addition to experiencing traffic surges due to demand that is synchronised in time, ISPs also experience instances of traffic demand that create network ``hot spots'', i.e.,~overloaded network links or servers. To mitigate congestion in these circumstances, DSM could encourage users to interact with topologically-distinct regions of the network. For example, the ISP might ask users to use augmenting services (e.g.,~peer-to-peer mechanisms) to allow ISPs to redirect and manage web requests. Alternatively, an ISP might ask a user to prefer services for which the ISP has more ample provisioning. For example, some services mirror content across multiple platforms. Based on content delivery network (CDN) provisions, ISPs might have different preferences for which service their users connect to. This could also occur for individual CDNs, with the ISP expressing a preference for which CDN node a client is redirected to (either by interacting directly with the CDN or, controversially, by manipulating the CDN's DNS responses).


In more radical settings, future ISPs could also exist within a far more fluid marketplace where user traffic could be shifted dynamically between entirely different ISPs on a minute-by-minute basis. Such an open environment could allow ISPs and users to negotiate which ISP would handle each network interaction. In such a case, a user may check their emails via one ISP, and visit a website via another.

\three~{\bf Content shifting}. In some cases, it may be possible to shift users' preferences in terms of \emph{what} they use the Internet for (e.g.,~which content they consume). For example, content curation and recommender services could encourage subscription to ``programs like $x$'' as a way of influencing user choices to become more easily cacheable --- using the ``filter bubble'' effect to the benefit of the ISP. Such features could also be used to shape different users towards different caching infrastructures within the ISP itself (or its peering points) in an attempt to load balance. When this type of filtering is applied to large populations of users, these approaches could have a significant effect on aggregate traffic. In extreme cases, users might even be offered alternative content (e.g.,~nearby cached items) at request-time to mitigate peak traffic loads. This latter point seems to have strong potential considering the power of recommendations~\cite{Davidson:2010} and users' consumption flexibility exhibited for some types of content~\cite{Tyson:2013}. Note that this is quite different to traditional content manipulation, which is generally limited to bit rate variations~\cite{thang2012adaptive}. We argue that in many cases users would \emph{rather} watch a different video at a high rate than their ``ideal'' video at a low quality.


\subsection{User Incentives}

All of these forms of shifting require cooperative adaptation of user behaviour. Thus, there is a basic need for robust incentives that communicate the benefits of behavioural change to the user. These could, for example, be presented to the user via a browser plug-in. We consider four of the many possible models, which are applicable to both mobile and wireline environments:

\one~{\bf Pricing}. One well-established way to incentivise behavioural change is through pricing. Pricing has typically been investigated to implement timeshifting strategies~\cite{Sen12}, but other activities (e.g.,~participation in network-friendly or ISP-local peer-to-peer file-sharing) could also be directly rewarded with discounts on contracts, or by metering different traffic types in different ways. More complex inter-organisation pricing arrangements might also be constructed~\cite{icqt,mortier02:multi.inter}. For example, ISPs with so-called ``usage caps'' might encourage users to adjust their behaviour by only counting some types of traffic against the usage cap. This practice is sometimes called unmetering, or zero-rating, as recently promoted by Facebook. ISPs in countries such as Australia have already implemented unmetering on a limited basis for certain content~\cite{www-whistleout-unmetered}. British Telecom also do this, allowing unmetred access to the iPlayer video on demand service from YouView set-top boxes.


An ISP could also reduce prices for other services such as Netflix, offering (free) prioritised delivery for those providers in return. Such prioritisation may actually incentivise customers to join the ISP if they frequently use the prioritised content providers. In markets where users have a choice between multiple ISPs, dynamic pricing might also be coupled with flexible switching between ISPs, allowing users to move between virtual ISPs on shorter timescales. ISPs might even transparently exchange customers to adapt to their operating conditions.

\two~{\bf Sociological}. Another possible incentive model could rely on users' willingness to adapt to achieve wider societal benefits. For example, a prominent societal concern is the increasing energy footprint of IT infrastructure. Users' behaviour might be affected by making such impacts explicit to them while they are using the Internet (e.g.,~by presenting the predicted energy consumption required to refresh a webpage). Past work has found that communicating network information to users~\cite{chetty2011my} is actually something appreciated. This information could be integrated into their browsing experience (e.g.,~via a browser plugin) to encourage users to avoid unnecessary network usage. Many variations of this notion exist, including making users aware of the congestion they're creating and its impact on others in the network neighbourhood. Such features could perhaps even be integrated into existing social networks to notify users when their activity impacts their friends.

\three~{\bf Service provisioning}. Another reward for behavioural change could be through the alteration of service and network provisioning for cooperative users (our earlier measurement studies have highlighted that variable service provisioning is already widely practised, though not always explicitly). For example, users who timeshift consumption by an hour might be guaranteed a higher quality of service, or perhaps ad-free content. Such rewards could become explicit service offerings, such as the ``nighttime data bundles'' that ISPs in certain countries offer~\cite{www-mweb-packages}. Alternatively, users could accumulate credits enabling better service provision in the future. This type of ``loyalty reward'' has proven extremely successful in other markets, such as airlines and supermarkets.

\four~{\bf Preference shaping}. A final alternative would be to personalise content offerings for individual users by presenting them with different options (e.g., different lists of content). Recent work~\cite{Tyson:2013} found that users in certain domains are often flexible about the content they consume (and are heavily impacted by browsing order). In these situations, ISPs could encourage content shifting by presenting options to users in different orders; for example, locally cached content could be placed at the top of a recommendation or search-result list. Such functionality could be attained by forming a dialogue between content providers (e.g.,~YouTube) and ISPs. For example, an ISP could ask that YouTube adapts its front-page and recommendations for users from their network by, for example, promoting locally cached content in recommendations. In return, YouTube might ask ISPs to prioritise traffic for selected premium content. In other situations, the ISP might also be the content provider making this dialogue even easier (e.g.,~Comcast/NBC in the USA is both an ISP and a content provider; BSkyB in the UK is in a similar position).

Controversially, this approach could allow ISPs to implement DSM without user awareness (indeed, they may already be doing so). We argue that, in the spirit of Human-Data Interaction~\cite{hdi}, these practices should be made explicit in the contracts between ISPs and users, and between ISPs and content providers. Further, users should be provided with mechanisms to opt-out at any time. Understanding the broader ramifications of such mechanisms is a rich area for interdisciplinary research, with computer science, economics, policy and ethics all having a part to play. Whatever approach taken, though, we argue strongly that transparency should be a key requirement.

\section{Staggercast: DSM for ISPs}
\label{sec:managing}

Realising user behaviour adaptation is challenging. In this section, we propose a strawman design, \emph{Staggercast}, named for its goal of staggering user demand over time, space and content. Again, we focus on video as a use case because it is such a significant fraction of Internet traffic.

In Staggercast, network interactions (e.g. a \textsf{GET} request) falling under DSM would be directed via a \emph{DSM Proxy} in an edge network. The remit of this middlebox would be defined by the ISP. For example, if the ISP wishes to manage Netflix usage, all Netflix traffic would be diverted through a DSM proxy. The proxy would then apply a DSM rule-set to decide whether the request should be subject to management procedures. This decision might be made on many factors, including \one~current traffic loads; \two~transit pricing; \three~user contract and past behaviour; \four~download size; and \five~quality-of-service constraints. 

If the proxy decides to perform DSM, the flow will be routed to the ISP's \emph{DSM Enactor}, which would implement one or more strategies that attempt to modify the users' behaviour. Clearly, these components exhibit synergies with software defined networking (SDN), offering an existing platform upon which Staggercast could be built. Based on the mechanisms discussed in~\S\ref{sec:dsm}, we provide three examples of how DSM strategies could be realised:

A user's request could be paused, and a ``staging'' webpage returned instead, informing the user of the ISP's desire to alter their behaviour. A variation of this would be to ask that users install a daemon or browser plug-in allowing such information to be conveyed in a more seamless manner. This has proven effective in related areas of traffic management~\cite{chetty2015ucap}. For example, the webpage or plug-in might ask the user to to delay their request or install a P2P application that matches users with peers within the ISP. Alternatively, it might simply inform the user of the negative impact their behaviour is having on others. Of course, it would also communicate any incentives available, i.e.,~the rewards for cooperation. The user would then be offered the choice of adapting their behaviour or continuing as normal.

A user's request could be transparently adapted or, alternatively, the response could be changed without user awareness (assuming prior and continuing consent). For example, an HTML response could be re-written to place nearby cached objects higher up in a browsing order. This would require deep packet inspection or application-layer proxies. Alternatively, it could be performed directly by the content provider via (automated) negotiation with the ISP.

A user's request could be optimised somehow, e.g.,~redirected to a nearby cache or alternate provider that is not congested. This is a form of transparent intervention, but would not directly change user behaviour. It would, however, potentially rely on prior user actions such as the installation of a P2P application (so that the HTTP request to the site can be transformed into a P2P one). Hence, in many cases these strategies must be used in tandem to realise DSM.

Some of these techniques might be progressively deployed without requiring client-side modification. That said, to better enable the above functionality (e.g.,~automated installations of software, pre-fetching of content), software could be embedded within the home access point that could mediate interactions between the user and the DSM proxy so that selected actions could be taken automatically. To reduce deployment costs, such software could also be installed in other nearby locations such as the Broadband Remote Access Server. Importantly, by exploiting these SDN-like principles, novel strategies could be implemented and dynamically introduced without significant alteration to existing infrastructure. As previously stated, all of these mechanisms must be underpinned by appropriate user consent, posing an interesting challenge for human-computer interaction.

\section{Preliminary Exploration}
\label{sec:survey}

The key to success in Staggercast is stakeholder cooperation. This is likely the largest hurdle as, generally speaking, Internet users in the developed world are accustomed to low cost flat pricing. As a first step in exploring this issue we distributed a questionnaire via social networks and three technical research group mailing lists to Internet users asking how they would deal with having DSM applied to their Internet use. We received 100 responses, primarily from the UK and the USA.\footnote{Space constraints restrict further demographic investigation, but the survey and data are available online, \url{http://www.eecs.qmul.ac.uk/\~tysong/files/staggercast_survey.pdf}.} The respondents were relatively heavy Internet users: 87\% said that they spent in excess of 3 hours online per day, with 69\% saying that the Internet is indispensable to them. Further, 43\% report that they watch online video more than 3 days a week, suggesting they have heavy traffic profiles. That said, we wish to emphasise that we do not use this survey as scientific evidence but, instead, as a mechanism to drive discussion. Our next stage of work would be to expand this work to involve field trials and ethnography.



\subsection{Timeshifting user activity}

\begin{figure}[tb]
 \includegraphics[width=80mm]{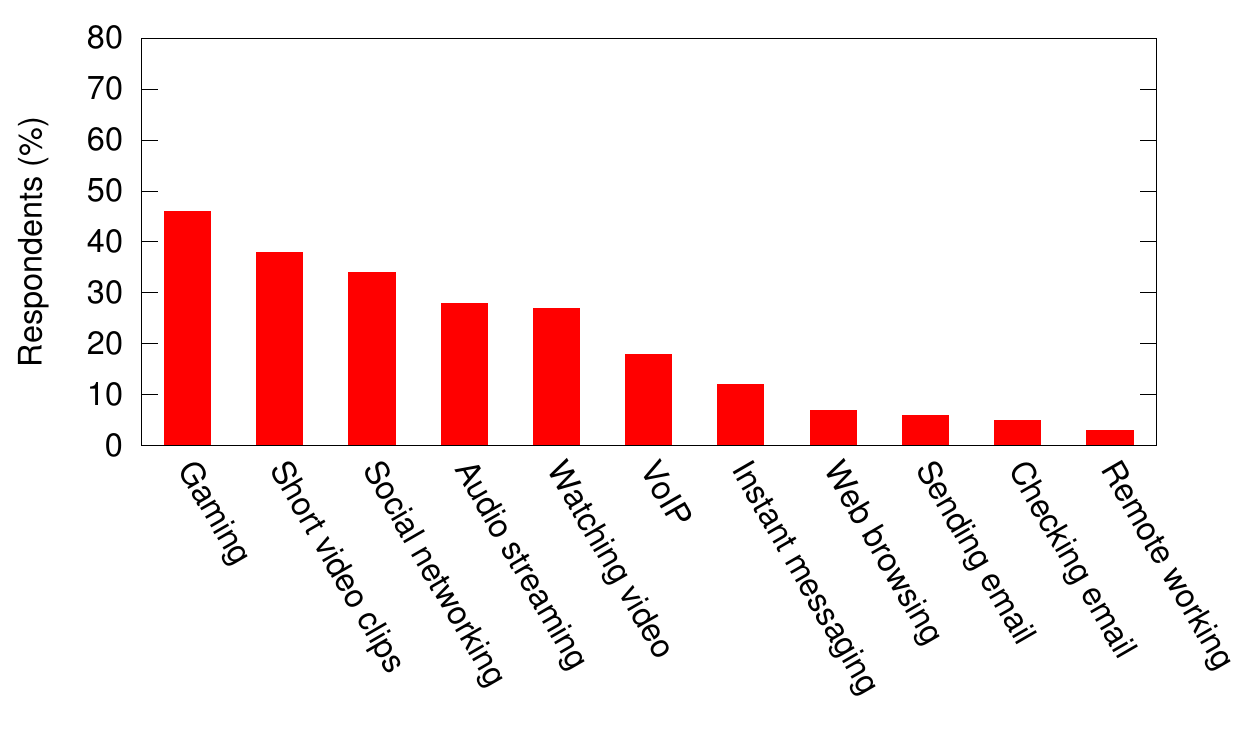}
 \caption{Percentage of respondents who would delay using an Internet application for over 3 hours}
 \label{fig:delay_3hrs}
\end{figure}

We asked users if they would be prepared to timeshift Internet usage given one of several incentives. Only 28\% said no, with 70\% saying that they would do it either often or occasionally.\footnote{Note that not all respondents answered all questions, marking some as N/A.} We further enquired about the applications they would be prepared to timeshift on. Figure~\ref{fig:delay_3hrs} presents the percentage of respondents who said they would be prepared to timeshift for more than 3 hours (per application). It can be seen that they feel some applications, such as gaming and video streaming, are quite flexible in terms of when they are used. In contrast, other applications were considered less flexible, with only 6\% of respondents saying they would timeshift sending emails for longer than 3 hours, for example. Broadly speaking, it seems that our respondents are more flexible about entertainment-oriented applications, with work-based applications being far less flexible (54\% of people said they would \emph{never} timeshift remote working). Importantly, the nature of peak time  operation and the relative weight of usage involved means that you only need to persuade a small number of customers to timeshift to have a significant impact (e.g., shifting 5\% of traffic can have a notable calming impact on peaks).


\subsection{Incentivising change}

\begin{figure}[tb]
 \includegraphics[width=80mm]{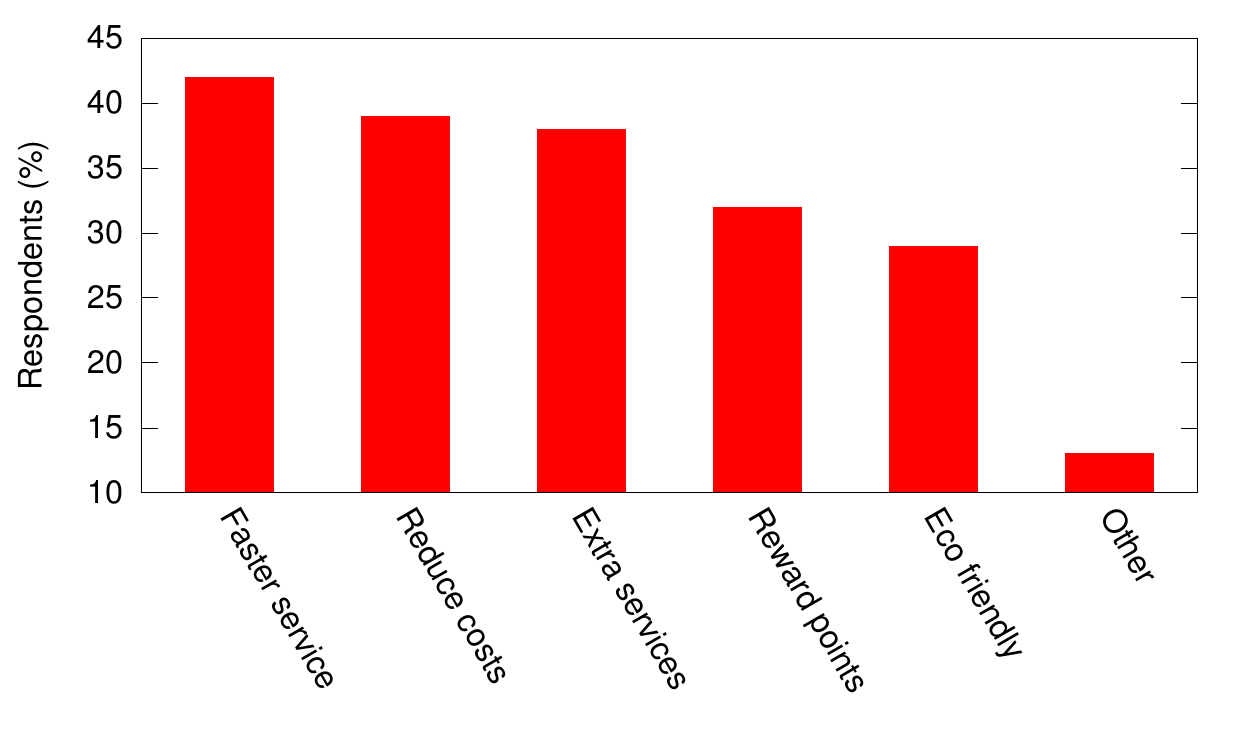}
 \caption{Percentage of respondents who stated that an (indpendent) incentive would encourage them to delay using Internet applications}
 \label{fig:incentives}
\end{figure}

Many users in the survey said that their responses would be based on the particular incentives offered. Thus, we asked questions about which incentives would drive them to shift behaviour, shown in Figure~\ref{fig:incentives}. The most popular incentive was guaranteeing the users a high quality of service later on, with the second two most popular relating to pricing: cheaper contracts or discounted extra services such as Netflix. We were also surprised to see that certain less conventional incentives were popular with the respondents: 29\% said they would be encouraged to change behaviour if they thought it would be more environmentally friendly. In fact, one respondent specifically highlighted (in free-text) that they would \emph{not} be prepared to engage in any incentives other than environmental ones. This highlights the need for personalisation in incentive models.

We also asked the users to propose their own incentive schemes. In more advanced DSM, this could even be performed as part of a negotiation between customers and network providers (where users petition ISPs for new incentives). Most respondents (76\%) did not state anything here and many actually stipulated that no incentives were appropriate, due to the way they use the Internet. One reason given for this was simply that they perceived the Internet to be so cheap that it would never be necessary to shift. Some respondents, however, did propose incentives. One asked that the shifted traffic did not count towards their data usage, whereas another thought that superior auditing of their traffic would be sufficient (e.g., offering a breakdown of their usage, alongside documented benefits they have received). This raises an interesting side-effect: Any sort of DSM \emph{must} be highly transparent, effectively presenting incentives and implications to the customer (particularly vital considering ongoing furore surrounding net neutrality issues). 



\subsection{Video shifting}

Network providers often complain of the difficulties in supporting video in an affordable and scalable way. We therefore dedicated a section of the survey to people's willingness to change their video consumption patterns. Our survey base were heavy video users, with 17\% saying that they used video on demand everyday and 43\% reporting that they used it more than 3 days a week. 51\% of respondents said they usually sit down with a specific video in mind, making content shifting difficult. However, the remaining participants said they were less rigid. 8\% of people even said they usually watch whatever is highlighted on the front page.

\begin{figure}[tb]
 \includegraphics[width=80mm]{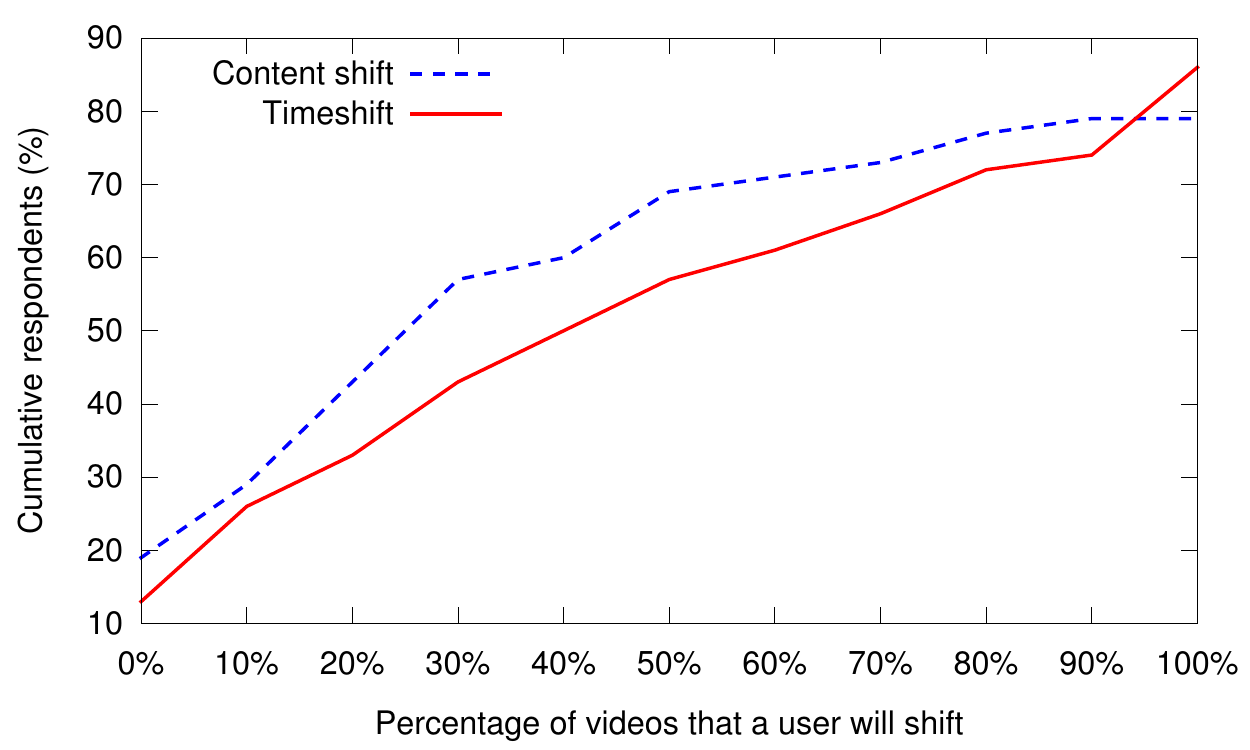}
 \caption{\label{fig:pc_flexible_vids}Cumulative percentage of videos for which respondents were prepared to consider substituting an alternative video}
\end{figure}

We asked if users would be prepared to watch an alternative video if offered (i.e., content shifting). 30\% said never, 56\% said occasionally and 6\% said often. We asked participants what percentage of videos they were prepared to timeshift and content shift. Figure~\ref{fig:pc_flexible_vids} presents the cumulative number of respondents for each percentage point. Overall, it can be seen that our respondents are more willing to timeshift than content shift. That said, 19\% said that they would be willing to consider watching an alternative for 50\% or more of the videos that they sit down to watch. The key criteria for watching the alternative listed were how much the consumer wanted to watch the original video vs. the alternative proposed video. Interestingly, the actual incentive offered did not appear to be a prominent factor in this decision with only 8\% of people ranking it as highly important. We were also surprised to see that the respondents were relatively flexible in \emph{when} they watch their content. 15\% said they would be \emph{always} prepared to delay watching content, and only 12\% said they would never delay.

Another interesting observation is that their willingness to content shift varies based on the content genre. For example, 44\% said that they would be very unlikely to delay the watching of sport, whereas this was only 19\% for movies. Clearly, such factors must be taken into account when making decisions in content shifting. This relates to the personalisation of incentives previously mentioned. A per-user approach must be taken if DSM is to succeed.

Finally, we asked users which incentives would work best for video-on-demand specifically. The results here were different from the trends discussed earlier. The most popular incentive was to remove advertisements from video streams with 52\% saying this would encourage them the most to shift. 30\% also said being rewarded with superior video quality would be acceptable too. This is something that should be noted by providers, as the usual response to congestion is the downgrading of content quality (via MPEG-DASH). It seems many users maybe willing to alleviate congestion by delaying consumption instead. 

\section{Related Work}
\label{sec:related}

Demand-side management has its roots in power networks: energy companies have long used differential pricing to encourage users to shift usage to different times of day, e.g.,~the UK energy companies' \emph{Economy 7} tariff. Odlyzko suggested in the late 1990s that this could be applied to the Internet~\cite{pmp-odlyzko}. Recently, Prabhakar {\em et al.} have applied similar concepts to road networks, using incentives to encourage commuters to defer travel to off-peak hours~\cite{prabhakar2009:netecon}.

Staggercast builds on previously proposed notions of bandwidth sharing and time-shifting traffic demands.  Researchers have so far focussed on various forms of time-dependent pricing~\cite{sen2013price}. Chhabra {\em et al.} proposed using incentives based on existing flat-rate schemes that allow homes to select a portion of their peak utilisation and move it to off-peak hours; they focus on shifting delay-tolerant traffic (e.g., peer-to-peer) in exchange for higher download rates during off-peak hours~\cite{chhabra2010:homenets,laoutaris2008:hotnets}. The Trade and Cap system offers an incentive mechanism for allowing users to trade bandwidth allocations at a shared link~\cite{londono2010:netecon}; the INDEX project studied how home network users adjust usage in response to different pricing structures~\cite{varian2002:demand}. Other work has focused on pricing schemes for shifting mobile traffic \cite{ha2012tube},  as well as for cloud resources~\cite{Xiang14}. We argue that this work is too narrow in scope, relying on pricing schemes to mediate resources. Hence, we build on this work by introducing new shifting types (content and space), alongside moving the focus away from price-based incentives and towards more sophisticated incentive mechanisms. Our key goal has therefore been to offer a generalised and extensible framework that can be used by ISPs to encompass all these forms of incentive-driven behavioural changes. To the best of our knowledge, we are the first to propose such a framework.



\section{Conclusion and Agenda}
\label{sec:summary}

The continuing growth in demand for online content from home access networks threatens to outpace the capacity provisioned at the edge, particularly as users increasingly request ``over the top'' video  streaming during peak hours. Although ISPs could handle this increasing load by provisioning extra capacity, doing so in the access network remains expensive and cannot adapt sufficiently quickly to respond to rapidly shifting demands. In this paper, we have proposed an architecture for {\em Internet Demand-Side Management}. We have outlined several techniques that ISPs might employ to ``manage'' the behaviour of users, and a preliminary architecture, Staggercast, which could implement these strategies.

There are many avenues of exploration remaining. Implementing and deploying Staggercast in a realistic setting is an important challenge. Such a system must determine which traffic demands should be shifted and what incentives should be offered to which users to shift demand appropriately. Hence, the coupling of caching/storage decisions and the use of incentives to affect user behaviour is a ripe area for further explanation. Effective DSM also requires a better understanding of the effectiveness of incentives on user behaviour through deployment and user studies. In particular, we must determine how different users respond to different incentives, measuring elasticity of traffic demand in practice. In the case of video shifting, there is also research to be done in predicting \emph{which} alternate videos might be acceptable to a given user, and under \emph{what} incentive.

The architecture also potentially creates new markets to design, explore, and study. For instance, ISPs might sell storage space at the edge to content providers, who can then encourage users to shift content viewing patterns to make more effective use of caching, e.g.,~reducing advertising in free content, or offering paid content at a lower price. Undoubtedly, our proposal raises many questions and gives rise to potential concerns regarding net neutrality, particularly when considering strategies involving content shifting and manipulation of recommendations to affect demand. In some cases, the user might even be completely unaware that they are participating in DSM, which also raises legal and ethical questions relating to disclosure.

Staggercast must therefore take a position in the debate; some may argue that this position is against net neutrality. However, Staggercast aims at enabling greater transparency and flexibility in the relationship between ISPs, content providers and consumers. In any case, unless regulation is enacted to prevent such behaviours, it appears such trends are inevitable, making it imperative that we explore more flexible ways to manage content and traffic, lest we miss an opportunity to improve network efficiency. In some cases, users may, in fact, benefit by paying for (and receiving compensation for) services that more accurately reflect the value of delivering bits over the network. Ultimately, by making the costs of delivering content more transparent through flexible incentives, the efficiency of network markets may improve. As one example, ISPs well-provisioned enough to have excess capacity at peak times might \emph{encourage} their users to make content requests. Such excess capacity could be used to fill local caches, which users of \emph{other} ISPs could pay to access (e.g.,~by creating a ``spot market'' for caches). Thus, in addition relieving peak-time congestion, bandwidth scarcity in the access network can be turned into a source of profit for both access ISPs and users.

We acknowledge that such an approach is inherently risky: it puts network neutrality to the test; it has considerable security, privacy, and ethical implications; and users themselves may not participate in the incentive mechanisms if the they are not made sufficiently transparent and easy to understand. Yet, we believe that these challenges -- and the potential positive benefits of a successful DSM architecture -- are precisely what makes this research area worth exploring further.

\end{sloppypar}

\begin{small}
  \bibliographystyle{abbrv}
  \balance\bibliography{Bibliography}
\end{small}

\label{lastpage}
\end{document}